# Perceptions of Smartphone Users' Acceptance and Adoption of Mobile Commerce (MC): The Case of Jordan

[1]**Ahmad Nabot,** [2]**Firas Omar** and [3]**Mohammed Almousa**

[1]*Department of Software Engineering, Zarqa University, Zarqa, Jordan*
[2]*E-business and Commerce, Petra University, Amman, Jordan*
[3]*Department of Software Engineering, Zarqa University, Zarqa, Jordan*

**Abstract:** This study investigates smartphone users' perceptions of adopting and accepting Mobile Commerce (MC) based on users' perceived adoption under the extended Technology Acceptance Model (TAM2) and Innovation Diffusion Theory (IDT) by providing research constructs for the domain of MC. Also, testing them with reliability and validity and demonstrating their distinctiveness with hypothesis testing. The results show that consumer intention to adopt MC on a smartphone was primarily influenced by Uncertainty Avoidance (UA), User Experience (UX), Perceived Ease Of Use (PEOU), Perceived Usefulness (PU) and Compatibility (CMP) as well as other constructs that positively determine attitude toward using a smartphone. For researchers, this study shows the benefits of adapting TAM constructs into MC acceptance on a smartphone. The perceptions of MC adoption on a smartphone in this study investigated based on a survey of specific people. For more reliability, a comprehensive study is needed to show the attitudes of people from different environments.

## Introduction

Recent and rapid developments in modern wireless communication technologies have led to a high rate of Internet penetration among smartphone users. Thus, Mobile Commerce (MC) has become increasingly significant for both enterprises and consumers Pascoe *et al*. (2002) Rupp and Smith (2002). Besides, the appearance of broadband ten years ago has replaced dial-up Internet connection, which became the primary Internet means of access for one billion users during that period Brown (2015). After that, the new generations of wireless networks (e.g., 3G and 4G) started replacing the older versions of these networks. According to Internet society. Org and smart insights.com statistics Brown (2015) Chaffey (2018).

In 2015, most of the world's countries had 3G mobile networks that covered 50% of the global population, where the number of Internet users reached 3 billion. Also, Internet usage on a smartphone is forecasted to be 71% by 2019 and the usage per device is forecasted to be more than triple in the same period. Thus, revenues from global online trade will increase, where over $ 230 billion will be revenues from MC Sharrard *et al*. (2001) Wu and Hisa (2008).

However, insufficient user acceptance of adopting new Information Technology (IT) will be a hurdle for the development of such technologies, specifically, with the rapid and extensive developments in mobile technology and MC applications. Therefore, there is a crucial need to understand MC consumer perceptions and acceptance of such technology. MC presents many advantages to its users, such as self-efficacy, convenience, a broader selection of products and sellers, competitive prices and products' rich information. Consequently, developments of e-commerce and internet services, including advertising, shopping, investing, banking and other online services have made it possible for people to change their daily lifestyles through interacting with the Internet. Also, with accelerated business competitions and the spread of Internet and smartphone usage, there is a need to understand the factors that would attract users to use MC. To investigate perceptions of smartphone users' acceptance of using the new technology, a large number of articles used the Technology Acceptance Model (TAM) and Innovation Diffusion Theory (IDT) as basic models. While a few studies investigated smartphone users' perceptions of MC and other factors affecting their perceptions, including User Experience (UX) and Uncertainty Avoidance (UA). In this study, the extended Technology Acceptance Model (TAM2) and Innovation Diffusion Theory (IDT) will be used to determine factors that affect consumers' perception of adopting and accepting the use of MC. Because of the rapid diffusion of Information Technology (IT) around the globe, theoretical research of smartphone and PDA devices adoption investigates the perceptions of its users. This study contributes to the information technology research field, with the diffusion use of smartphone technology adoption, by explaining factors affecting users' perceptions in adopting such technology and MC. This study aims to investigate human motivations affecting their perceptions of adopting and accepting Mobile Commerce (MC) as a smartphone and PDA applications. The study will provide more in-depth insight to identify the factors that affect consumers' decisions to adopt MC on smartphones by employing TAM2 and IDT as basic models. Also, a hypothesis of the individual attitude to adopt MC on smartphones is determined by perceived ease of use, perceived usefulness, compatibility, effectiveness, efficiency and other factors that affect their decisions. Eventually, the proposed model in this study will help to understand the influencing factors of smartphone and PDA users' perceptions and provide futuristic research suggestions and developments in this scope.

This study begins with two key goals: (1) Reviewing the available literature on user intention toward adopting MC and (2) understand TAM2 and IDT constructs to see the most influencing factors to extend TAM2 by adding new factors. The remainder of the paper is structured as follows: First, a review of the existing literature on investigating user perceptions and attitudes toward adopting MC and presenting the study hypothesis. Next, study methodology and affecting factors discussed and after that, presenting study results of the collected and analyzed data. Finally, discussions and conclusions of the study presented.

## Main Concepts, Research Hypothesis and Model

Mobile Commerce (MC) refers to monetary transactions implemented via an Internet connection using smartphone technology Barnes (2002) GANDHI (2016). Therefore, vendors, service providers, information systems and application developers must guardedly understand the various needs of smartphone users to provide high-quality services that entice them to adopt MC Wu and Wang (2005). MC considered a kind of e-commerce that has many types, where the most used types are B2C and C2B that depend on the wireless network to complete transactions. These transactions include shopping, browsing, online payment transactions, etc. Eastin (2002). Nevertheless, MC is considered to be the future of banking services where most people will use this new technology to complete their transactions through smartphones because of many reasons, i.e., convenience, secure transactions, cost-effective offerings and the ability to complete transactions from anywhere.

### *TAM2 and IDT*

The Technology Acceptance Model (TAM) was introduced by Davis (1989) and used for investigating and predicting users' behavior toward adopting the use of information technology Rupp and Smith (2002). The model derived from the Theory of Reasoned Action (TRA), which considered as a base of TAM Wu and Wang (2005). Since its development, TAM consisted of two main factors to determine users' intentions to adopt and accept technology. These factors lie in Perceived Ease of Use (PEU) and Perceived Usefulness (PU) Gao (2005). Venkatesh and Davis (2000) extended the TAM model into the TAM2 version to include additional factors due to the different behaviors and environments of the users. Additionally, the spread of a wide range of applications and its usage in different fields of life such as health, engineering, entertainment, etc. has led researchers to extend this model more and more Legris *et al*. (2003) Hamid Shokery *et al*. (2016). The extended TAM or "TAM2" included extra factors to predict users' intentions to accept and adopt the use of information technology, such as subjective norms, hedonic, utilitarian factors, etc. Balog and Pribeanu (2016) Kim *et al*. (2017). There is a relation between TAM and Innovation Diffusion Theory (IDT) in terms of the constructs' dependability, whereas the constructs of one model supplement the constructs of the other Sánchez-Prieto *et al*. (2016). The main idea of innovation diffusion is "the process by which an innovation is communicated through certain channels over time among the members of a social system" Agag and El-Masry (2016). IDT presents several constructs that play an essential role in influencing users' decisions to adopt new technologies. These constructs are relative advantages such as compatibility, complexity and trial-ability and visibility Venkatesh and Davis (2000). Since the purpose of this study was to understand users' thoughts, concerns and experience concerning the adoption of MC, TAM was found to be the most suitable model to investigate and explain users' attitude toward adopting new technologies due to its reliability and validity Rupp and Smith (2002). After that, the smartphone and its related applications started to appear and spread quickly among users, stimulating researchers to investigate consumers' attitudes toward adopting the new technology. Therefore, this study comes to explore consumers' attitudes toward adopting and accepting MC technology. Perceived usefulness, perceived ease of use and behavioral intention as the main factors of TAM considered as the essential factors for influencing consumers' decisions for adopting this new technology. Therefore, the following hypotheses formulated:

H1. Perceived usefulness positively affects behavioral intention
H2a. Perceived ease of use positively affects behavioral intention
H2b. Perceived ease of use positively affects the usefulness

According to Chen *et. al,* (2002) the compatibility construct of IDT could provide a further investigation of consumers' attitudes toward adopting MC when combined with the original TAM's behavioral intention constructs. Therefore, the following hypotheses have formulated:

H3a. Compatibility positively affects behavioral intention
H3b. Compatibility positively affects the usefulness

Continuously, the new factors proposed by this study are User Experience (UX) and Uncertainty Avoidance (UA), which all influence consumers' intention to adopt MC and mobile services.

*User Experience (UX)*

User Experience (UX) is one of the most influencing factors affecting consumers' attitudes towards m-commerce adoption, where many firms try to use this factor to create a competitive advantage and excellent user experience Bilgihan *et al.* (2016). Therefore, according to Albert, W. and Tullis, T., a UX term is defined as "when a user is involved in interacting with a product, or system interface due to user interest in observing or measuring something." Thus, user behavior or attitude toward using technology considered as UX due to the user-ability to evaluate any system through interacting with its interface. Also, UX takes into consideration the users' entire interaction with the system or application through feelings, thoughts and perceptions as a result of the interaction William and Tullis (2013). However, UX is a crucial part of the development process of any new technology because it has a broader view of evaluating the product itself and the users' attitude in using such product through different metrics. These metrics are efficiency, effectiveness and user satisfaction, which considered as the critical factors in improving user experience Hokkanen *et al.* (2015). Also, UX metrics help to achieve a better understanding of the users' attitude toward adopting new technologies and even to detect severe inefficiencies in the product or system, which has a relation with some goals of Human-Computer Interaction (HCI) discipline William and Tullis (2013) Diaper and Sanger (2006). Each metric of UX metrics relates to a specific function aspect in the desktop and mobile applications and connectivity, which considered as an indicator of the users' adoption intention Zarmpou *et al.* (2012). For instance, efficiency aspects related to the mobile application (response time, connectivity speed and the amount of the provided services by the application), effectiveness (performance and quality of the provided service) and satisfaction (users' satisfaction degree when performing the task) William and Tullis (2013) Nielsen (1993). Therefore, the following hypotheses proposed. Eventually, Dholakia and Kshetri (2004) and Büyüközkan (2009) stated that several constructs affect MC users' adoption intention, which also considered as essential requirements for such users. For example, complete interface," anytime-and-anywhere" capability and any other technical aspects that could affect application work behavior:

H4a. Efficiency positively affects behavioral intention
H4b. Efficiency positively affects perceived usefulness
H5a. Effectiveness positively affects behavioral intention
H5b. Effectiveness positively affects perceived ease of use
H6. Does subjective satisfaction positively affect behavioral intention?
H7a. How many times have you shopped online using your mobile?
H7b. Depending on your mobile commerce experience, how do you rate this experience?

*Uncertainty Avoidance (UA)*

Uncertainty avoidance is a cultural dimension that measures the level of society's ambiguity, trust and experience in using a specific product. According to G. Hofstede, (1991), Uncertainty avoidance (UX) plays a crucial role in the adoption of new technologies; where people live in developed countries such as the UK and USA have a lower uncertainty avoidance than people living in developing countries such as India, China and Jordan. Hofstede (1991). This is due to the sizeable dependent use of new technology in the daily lifestyle of people, the high level of awareness among people and the available ICT infrastructure in the developed countries. These factors helped in increasing the quality of life by increasing the quality of software products that made uncertainty avoidance lower than developing countries. Also, uncertainty avoidance considered a measure of two essential factors for m-commerce; these factors are security and trust. These two moderating factors of uncertainty avoidance considered as the main determinants of consumers' decisions toward adopting m-commerce, whereas two-thirds of consumers do not buy online due to security reasons Kao (2009). Whereas trust considered a crucial player of uncertainty avoidance, which could affect consumers with high uncertainty avoidance decisions to adopt MC Baptista and Oliveira (2016) Choi (2018). Therefore, the following hypotheses proposed:

H8a: Uncertainty avoidance moderating factors affect perceived usefulness
H8b: Uncertainty avoidance moderating factors affect the behavioral intention of people with high uncertainty to adopting MC

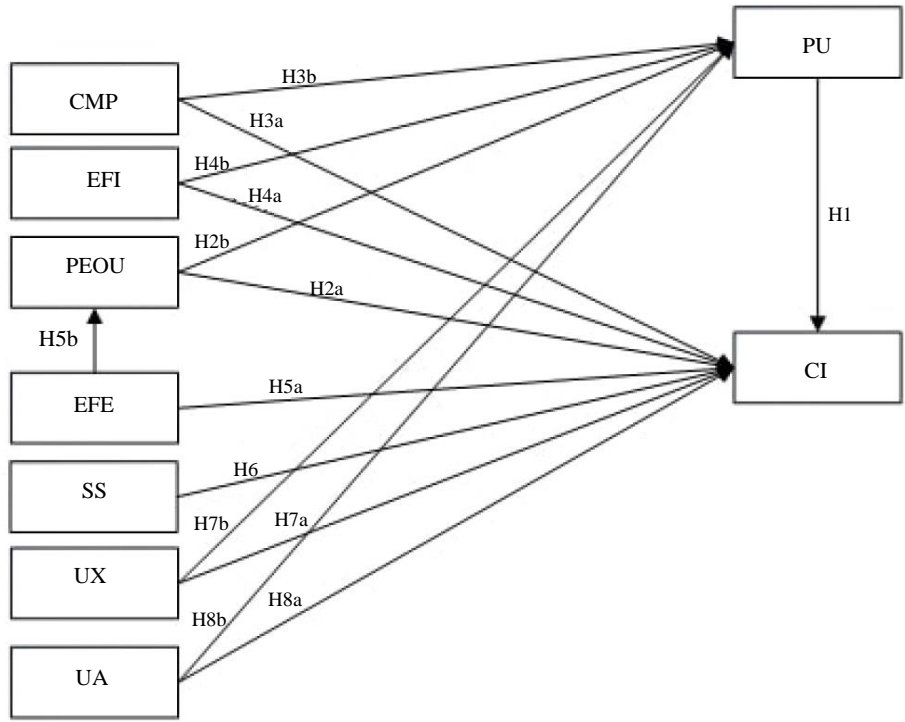

**Fig. 1:** The proposed research model for MC adoption

## Research Model

In this study, TAM2 with IDT has integrated with two additional factors (Uncertainty Avoidance (UA) and User Experience (UX)) that affect consumers' attitude to adopt MC. Other constructs, such as perceived ease of use, perceived usefulness, compatibility, efficiency, effectiveness and subjective satisfaction adopted from TAM2 and IDT. Figure 1 shows the primary factors of TAM2 and the new factors have integrated into this study.

## Research Methodology

### Instrument

The research hypotheses were empirically tested against the data collection using a survey questionnaire. The questionnaire was composed of constructs developed through iterative validation steps. First, based on extensive reviews of available research, the initial items of the constructs in the model were generated. Then, the initial version of the survey piloted by several people from the academic domain such as, university students and professors in Jordan. Finally, after collecting the feedback, some improvements took place in the survey and the constructs to better fit the research needs. Questionnaire items were measured based on a five-point Likert scale, ranging from 1 (Strongly disagree) to 5 (Strongly agree).

**Table 1:** Participants' demographic statistics

| Demographic profile | Frequency | Percentage |
|---|---|---|
| Gender | | |
| Male | 223 | 57.8 |
| Female | 163 | 42.2 |
| Age | | |
| 18-20 | 71 | 18.4 |
| 21-30 | 108 | 28.0 |
| 31-40 | 126 | 32.6 |
| Over 40 | 81 | 21.0 |
| Highest educational level | | |
| High school | 12 | 3.1 |
| Undergraduate\college | 228 | 59.1 |
| Doctoraten\Master | 146 | 37.8 |

### Sample and Procedure

A total of 500 questionnaires were distributed into different smartphone users who are expected to be using MC from the academic domain, such as university students and professors through email and Facebook groups. A total of 409 valid respondents obtained and 23 samples were incomplete and omitted from the analysis. Therefore, a total of 386 were considered to be valid for further analysis (response rate is around 85%). The high response rate was due to the convenient design of the questionnaire, which requires 10-15 min to complete. Of those 386 participants, 57.8% of them were males and 42.2% were females. The age of the participants ranged from 18 to over 40. About 32.6% of the participants were between 31 and 40 years old, 28% of the participants were between 21 and 30 years old, 21% of the participants were over 40 years old and 18.4% of the participants were between 18 and 20 years old. Finally, the educational level of the participants was ranging from high school to doctorate or master degree and most of the participants' educational level was 59.1% undergraduate, 37.8% doctorate or master degree and 3.1% high school (respectively), as shown in Table 1.

## Results

Questionnaire reliability was tested using Cronbach's coefficient alpha ($\alpha$) to estimate the internal consistency of all items that make up the scale Pallant (2013). Cronbach's alpha shall be 0.7 or higher for questionnaire items to be considered acceptable. Therefore, all questionnaire items tested and the reliability coefficient for all independent variables are above 0.7, confirming that all the items used to measure the constructs are reliable, as shown in Table 2.

The research model constructs at a cut-of-point of 0.5. A correlation matrix generated for all questionnaire items. Then, factors that have eigenvalues of more than 1.0 considered significant with factor loading of 0.5 as a cut-off point, while factors that have eigenvalues of less than 1 are considered insignificant and discarded Hair *et al.* (2006) Teo (2001) YeeâLoong Chong and Ooi (2008). Table 3 shows the first-factor analysis of the constructs, whilst, five items were removed from the analysis (EFI5, UA1, UA2, UA3 and UA5). EFI5, UA1, UA2 were loaded on a non-hypothesized factor and UA3 and UA5 were loaded on two factors instead of the hypothesized one. All other items had eigenvalues greater than 1.0 and factor loadings were greater than 0.5 on the factor hypothesized to load.

The second factor analysis was carried out using the remaining 22 items of the constructs to evaluate them after the first-factor analysis. Table 4 shows the rotated factor matrix of the second-factor analysis for the remaining items after the first-factor analysis that loaded on the proposed constructs. The factors in the analysis had eigenvalues greater than 1.0 and a total variance of 64.81 in the data.

## Hypothesis Testing

To test the study hypothesis, Multiple Linear Regression (MLR) analysis administered. Table 5 shows the results of the hypothesis testing (H1-H8) with P-value, a standardized coefficient ($\beta$) and a significance to test the relationships of the hypothesis in TAM. H1 test results indicated that PU had a significant positive impact on consumers' intention toward adopting MC ($\beta = 0.20$, $p < 0.000$). Hence, H1 is supported. For the tests of H2a, which indicated that PEOU had a significant positive impact on consumers' intention toward adopting MC ($\beta = 0.35$, $p < 0.000$). Hence H2a is supported. For the tests of H2b, which indicated that PEOU had a significant positive impact on PU ($\beta = 0.26$, $p < 0.000$). Hence, H2b is supported. For the tests of H3a, which indicated that CMP had a significant positive impact on consumers' intention toward adopting MC ($\beta = 0.16$, $p < 0.01$). Hence, H3a is supported. For the tests of H3b, which indicated that CMP does not have a significant impact on PU ($\beta = 0.010$, $p < 0.72$). Hence H3b is not supported. For the tests of H4a, which indicated that EFI does not have a significant impact on consumers' intention toward adopting MC ($\beta = 0.05$, $p < 0.29$). Hence, H4a is not supported. For the tests of H4b, which indicated that EFI had a significant positive impact on PU ($\beta = 0.14$, $p > 0.000$). Hence, H4b is supported. For the tests of H5a, which indicated that EFE does not have a significant impact on consumers' intention toward adopting MC ($\beta = 0.02$, $p < 0.51$). Hence, H5a is not supported. For the tests of H5b, which indicated that EFE had a significant positive effect on PEOU ($\beta = 0.23$, $p < 0.000$). Hence, H5b is supported. For the tests of H6, which indicated that SS does not have a significant impact on consumers' intention toward adopting MC ($\beta = 0.09$, $p < 0.23$). Hence, H6 is not supported. For the test of H7a and H7b, the regression outcomes of UX on CI towards adopting MC had a significant positive impact ($\beta = 0.46$, $p < 0.000$) and perceived usefulness ($\beta = 0.41$, $p < 0.000$). Hence, H8a and H8b are supported. Finally, for the tests of H8a and H8b, the regression outcomes of UA on CI towards adopting MC had a significant positive impact ($\beta = 0.33$, $p < 0.000$) and perceived usefulness ($\beta = 0.58$, $p < 0.000$). Hence, H7a and H7b are supported.

**Table 2:** Means, SD and cronbach's($\alpha$)

| Construct | Mean SD | Cronbach's ($\alpha$) |
|---|---|---|
| Perceived Usefulness (PU) | 3.12 1.61 | 0.912 |
| Perceived Ease Of Use (PEOU) | 2.37 1.76 | 0.897 |
| Compatibility (CMP) | 3.09 1.55 | 0.729 |
| Efficiency (EFI) | 3.72 1.16 | 0.732 |
| Effectiveness (EFE) | 2.96 1.22 | 0.752 |
| Subjective Satisfaction (SS) | 3.53 1.17 | 0.744 |
| Uncertainty Avoidance (UA) | 3.45 1.49 | 0.779 |
| User Experience (UX) | 3.34 1.31 | 0.764 |

**Table 3:** First-factor analysis

| Factor | F1 | F2 | F3 | F4 | F5 | F6 | F7 | F8 |
|---|---|---|---|---|---|---|---|---|
| PU1 | 0.838 | | | | | | | |
| PU2 | 0.848 | | | | | | | |
| PU3 | 0.836 | | | | | | | |
| PEOU1 | | 0.872 | | | | | | |
| PEOU2 | | 0.857 | | | | | | |
| PEOU3 | | 0.847 | | | | | | |
| PEOU4 | | 0.822 | | | | | | |
| CMP1 | | | 0.795 | | | | | |
| CMP2 | | | 0.779 | | | | | |
| CMP3 | | | 0.784 | | | | | |
| EFI1 | | | | 0.690 | | | | |
| EFI2 | | | | 0.744 | | | | |
| EFI3 | | | | 0.736 | | | | |
| EFI4 | | | | 0.675 | | | | |
| EFI5* | | | | | 0.381 | | | |
| EFE1 | | | | | 0.762 | | | |
| EFE2 | | | | | 0.768 | | | |
| SS1 | | | | | | 0.721 | | |
| SS2 | | | | | | 0.814 | | |
| UA1* | | | | | 0.568 | | | |
| UA2* | | | | | 0.317 | | | |
| UA3* | | 0.356 | | | | | | 0.383 |
| UA4 | | | | | | | 0.798 | |
| UA5* | | 0.619 | | | | | | 0.388 |
| UA6 | | | | | | | 0.818 | |
| UX1 | | | | | | | | 0.718 |
| UX2 | | | | | | | | 0.765 |

Notes: PU-perceived usefulness; PEOU-perceived ease of use; CMP-compatibility; EFI efficiency; EFE-effectiveness; SS-subjective satisfaction; UA-uncertainty avoidance; UX-user experience

**Table 4:** Second-factor analysis

| Factor | F1 | F2 | F3 | F4 | F5 | F6 | F7 | F8 |
|---|---|---|---|---|---|---|---|---|
| PU1 | 0.838 | | | | | | | |
| PU2 | 0.848 | | | | | | | |
| PU3 | 0.836 | | | | | | | |
| PEOU1 | | 0.872 | | | | | | |
| PEOU2 | | 0.857 | | | | | | |
| PEOU3 | | 0.847 | | | | | | |
| PEOU4 | | 0.822 | | | | | | |
| CMP1 | | | 0.795 | | | | | |
| CMP2 | | | 0.784 | | | | | |
| CMP3 | | | 70.779 | | | | | |
| EFI1 | | | | 0.690 | | | | |
| EFI2 | | | | 0.744 | | | | |
| EFI3 | | | | 0.736 | | | | |
| EFI4 | | | | 0.675 | | | | |
| EFE1 | | | | | 0.762 | | | |
| EFE2 | | | | | 0.768 | | | |
| SS1 | | | | | | 0.721 | | |
| SS2 | | | | | | 0.814 | | |
| UA4 | | | | | | | 0.798 | |
| UA6 | | | | | | | 0.818 | |
| UX1 | | | | | | | | 0.718 |
| UX2 | | | | | | | | 0.765 |

Notes: PU-perceived usefulness; PEOU-perceived ease of use; CMP-compatibility; EFI-efficiency; EFE-effectiveness; SS-subjective satisfaction; UA-uncertainty avoidance; UX-user experience

**Table 5:** Standardized path coefficients and P-value for the factors

| Hypothesis | Relationship | P-value | Standardized coefficients ($\beta$) | Result |
|---|---|---|---|---|
| H1 | PU⇒CI | 0.000 | 0.20 | Accepted |
| H2a | PEOU⇒CI | 0.000 | 0.35 | Accepted |
| H2b | PEOU⇒PU | 0.000 | 0.26 | Accepted |
| H3a | CMP⇒CI | 0.010 | 0.16 | Accepted |
| H3b | CMP⇒PU | 0.720 | 0.01 | Rejected |
| H4a | EFI⇒CI | 0.290 | 0.05 | Rejected |
| H4b | EFI⇒PU | 0.000 | 0.14 | Accepted |
| H5a | EFE⇒CI | 0.510 | 0.02 | Rejected |
| H5b | EFE⇒PEOU | 0.000 | 0.23 | Accepted |
| H6 | SS⇒CI | 0.230 | 0.09 | Rejected |
| H7a | UX⇒CI | 0.000 | 0.46 | Accepted |
| H7b | UX⇒PU | 0.000 | 0.41 | Accepted |
| H8a | UA⇒CI | 0.000 | 0.33 | Accepted |
| H8b | UA⇒PU | 0.000 | 0.58 | Accepted |

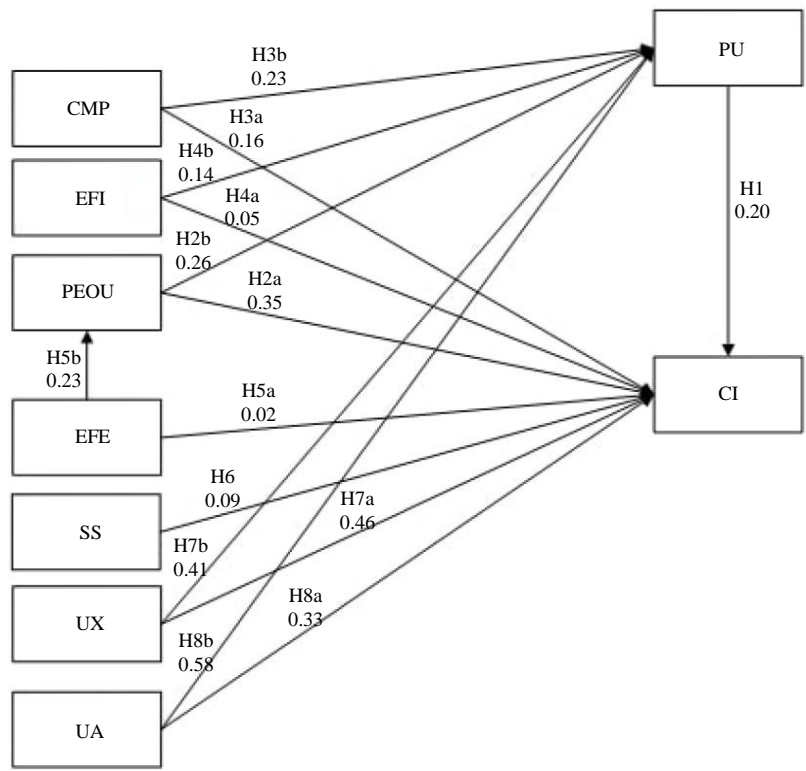

**Fig. 2:** Results of study research model

Figure 2 shows the main factors of TAM2 and the integrated factors with the standardized coefficients ($\beta$) of each factor after testing the study hypothesis using MLR. The results of MLR for each factor shows the significance of the relationship with the hypothesis in TAM.

## Discussion

Based on various theoretical studies, this study introduces a research model specifying key drivers of an individual's intention to adopt Mobile Commerce (MC) in their daily life activities. Using data from a large-scale survey conducted in Jordan, we found empirical support for the proposed model.

Test results in Table 5 indicated that uncertainty avoidance significantly affects customer intention and perceived usefulness, while perceived usefulness and perceived ease of use have a substantial effect on user intentions. Additionally, other factors used in this study such as compatibility, efficiency, effectiveness and subjective satisfaction, which also have moderate and weak effects on user intentions.

CI was positively affected by PU, which confirms the importance of these two factors. The findings also show a positive effect on CI from PEOU and a positive relationship between PU and PEOU as well; This implies that if smartphone users feel the easiness of using such technology and an improvement in their performance, then their intended outcomes will be improved towards using such technologies; This also confirms the compatibility of these results with previous studies of Adapa *et al*. (2017) Hubert *et al*. (2017) Yu *et al*. (2017).

Additionally, CMP has a positive impact on Consumers' Intentions (CI) toward using mobile commerce and a negative impact on PU. However, CMP considered an essential predictor of consumers' intention that plays a vital role in adopting such technologies. Therefore, MC managers should consider the needs, values and lifestyles of consumers that can be achieved through skipping compatibility issues to be more positively affecting consumers' intentions towards adopting MC, which is in line with previous studies of Agag and El-Masry (2016) Amaro and Duarte (2015) Wu and Wang (2005). Moreover, efficiency is considered as an individual's values and the efficiency of the used technology, such as software that saves time and money and enhances user experience Moorthy *et al*. (2017) Yu *et al*. (2017) Jan *et al*. (2019). In this study, we found that EFI had a negative impact on consumers' intention to adopt MC due to several barriers, such as the lack of developed infrastructure that hinder their intention, as well as the low level of awareness of the benefits of adopting such kind of technologies for shopping, especially in developing countries. Additionally, there was a positive impact from EFI on PU, which means that if the efficiency of the used technology improved, then customer values improve and their intention to adopt MC will improve as well. Also, mobile commerce offers convenience by offering a large number of products from different sellers and eliminating the need to travel for shopping, traffic, long checkout queues, etc., which is in line with previous studies of Basole (2004) Childers *et al*. (2001) Kim *et al*. (2009). However, Effectiveness (EFE) has a negative impact on consumers' intention to adopt MC and a positive impact on PEOU. Effectiveness and efficiency considered as dimensions of usability, which identified by ISO 924-11 to enable users to achieve their goals by using the complete product. Potentially, efficiency and effectiveness increase users' intention toward adopting MC by providing information services about products and product use to match customer needs. EFE was found to be a factor that has a positive impact on consumers' attitude in terms of PEOU. The findings are consistent with Basole (2004) Kim *et al*. (2009).

Subjective Satisfaction (SS) is identified as the degree of user satisfaction when using a product/service, which could be affected based on the provided service level or product quality Ström *et al*. (2014). Nevertheless, any deficiencies and incompleteness related to MC applications and services could negatively affect consumers' satisfaction. As shown in Table 5, users' subjective satisfaction has a positive impact on consumers' intention to adopt MC Dai and Palvi (2009). User Experience (UX) considered an essential factor that has a positive impact on CI and PU to adopt MC. Usability considered as an element of UX that influence consumers' intention toward MC Park *et al*. (2013). Zhou and Zhang (2007) suggested that user intention to make online purchases moderated by experience. However, UX has not been clearly identified due to its different aspects of interactions between the user, products and the provided services Alben (1996) Arhippainen and Tähti (2003) Forlizzi and Ford (2000) Kuniavsky (2007) Law *et al*. (2008) Law and van Schaik (2010) Marcus (2006) McNamara and Kirakowski (2006). While, Zabadi (2016) identified UX as the outcome that reflects the user's perception, the complete system characteristics and the context of use. Eventually, as shown in Table 5, UX has a positive impact on both CI and PU, which are related to each other and consistent with the findings of Zhou *et al*. (2007) Bendary and Al-Sahouly (2018).

Eventually, Uncertainty Avoidance (UA) also considered an essential factor that affects consumers' intention to adopt Dai and Palvi (2009) identified UA as "the degree of how societies accommodate high levels of uncertainty and ambiguity in the environment." Hofstede (1991) conducted interviews with several IBM employees in 50 countries to conduct the cultural dimensions that could affect cultures' intention to adopt MC. Also, previous studies took place to test the effect of UA on CI and found that UA has a positive impact on CI in developed countries' societies and a negative impact on CI in developing countries' societies Hofstede (1991). Also, UA considered as User Experience (UX) in technology usage and its related risk.

This study adopted the extended TAM model to show how consumers' intention related to MC acceptance among smartphone users. The findings showed that some of the most common factors such as compatibility, efficiency, effectiveness and subjective satisfaction had weak to moderate effect on the consumers' intention. These findings were consistent with the findings of Eneizan *et al*. (2016) Hong *et al*. (2008) Nassuora (2013) Kim *et al*. (2009) Alben (1996) Hubert *et al*. (2017). Also, Uncertainty Avoidance (UA), User Experience (UX), usefulness and ease of use had a significant effect on the customer intention, which consistent with the findings of Chung (2019) Zabadi (2016) Ameen and Willis (2018). To researchers, this study shows the most common and other constructs of TAM that make up the model for smartphone users' intention to accept and adopt MC. Although users' intention under TAM have been previously investigated, this study extended prior research by providing constructs for the domain of MC, testing their reliability and validity. In addition, using a more in-depth analysis to come up with more refined results of the used constructs.

## Conclusion

Although, MC is a new technology in some industries, thus, adoption of such technology deserves further investigations. This study contributes to the field literature by adding a new important investigation. Furthermore, it contributes to the literature by enriching it with an overview from Jordanian consumers' perceptions for adopting MC. The results of the previous studies are limited in the context of Jordan in comparison to other studies in developed countries. Therefore, one of the important implications is that organizational factors become a significant predictor of users' intention toward MC. The findings imply that managements should pay attention to the adoption decision of new technologies. Moreover, an enhanced communication infrastructure and software design of mobile applications to enhance its functionality and usability are considered as the most challenges that face businesses. Additionally, users and businesses are anxious about other specifications of MC applications such as efficiency, compatibility, robustness and security. Which require comprehensive development for such applications. As well as the lack of governmental laws and global standards for MC application usage.

This study provided valuable insights into the factors affecting consumers' intention to adopt MC, it has some limitations. First, the cultural characteristics of Jordanians in terms of shopping habits, the fear of making online payments and English language proficiency could affect their intention to adopt MC. Second, MC and online shopping in Jordan is still in its infancy and MC applications are limited, which lower user experience and affect their intentions to use MC. Third, the collected samples of this study were from academic domain in Jordan, which limits the findings from other people and cultures. Therefore, subsequent studies are required to investigate the findings of this study from larger samples of people and different cultures. Fourth, the study sample was biased to academic field people, such as university students and professors, which may lead to inaccurate results and perspectives. Eventually, Future research may investigate more constructs that have effects on consumers' intention to adopt MC from different cultures, which might yield rich and valuable insights.

## Acknowledgement

This research is funded by the Deanship of Research and Graduate Studies at Zarqa University /Jordan.

## Author's Contributions

All the authors contributed equally to this work.

## Ethics

This manuscript is the original contribution of the authors and is not published anywhere. There is no ethical issue involved in this manuscript.

Ahmad Nabot *et al*. / Journal of Computer Science 2020, 16 (4): 532.542
**DOI: 10.3844/jcssp.2020.532.542**Legris, P., J. Ingham and P. Collerette, 2003. Why do people use information technology? A critical review of the technology acceptance model. Inform. Manage., 40: 191-204.

Marcus, A., 2006. Cross-cultural user-experience design. Proceedings of the International Conference on Theory and Application of Diagrams, (TAD' 06), Springer, Berlin, Heidelberg, pp: 16-24.

McNamara, N. and J. Kirakowski, 2006. Functionality, usability and user experience: Three areas of concern, interactions.

Moorthy, K., C. Suet Ling, Y. Weng Fatt, C. Mun Yee and E.C. Ket Yin *et al.*, 2017. Barriers of mobile commerce adoption intention: Perceptions of generation X in Malaysia. J. Theoret. Applied Electr. Commerce Res., 12: 37-53.

Nassuora, A.B., 2013. Understanding factors affecting the adoption of m-commerce by consumers. J. Applied Sci., 13: 913-918.

Nielsen, J., 1993. Usability Engineering. 1st Edn., Elsevier Science, ISBN-10: 0125184069, pp: 362.

Pallant, J., 2013. SPSS Survival Manual. 1st Edn., McGraw-Hill Education (UK), pp: 354.

Park, J., S.H. Han, H.K. Kim, Y. Cho and W. Park, 2013. Developing elements of user experience for mobile phones and services: Survey, interview and observation approaches. Human Factors, 18: 95-111.

Pascoe, J., V. Sunderam, U. Varshney and R. Loader, 2002. Middleware enhancements for metropolitan area wireless internet access. Future Generat. Comput. Syst., 18: 721-735.

Rupp, W.T. and A.D. Smith, 2002. Mobile Commerce: New revenue machine or black hole? Bus. Horizons, 45: 26-29.

Sánchez-Prieto, J.C., S. Olmos-migueláñez and F.J. Garćıa-Peñalvo, 2016. Informal tools in formal contexts: Development of a model to assess the acceptance of mobile technologies among teachers. Comput. Human Behav., 55: 519-528.

Sharrard, J., S.J. Kafka and M.J. Tavilla, 2001. Global online trade will climb to 18% of sales. Bus. View Brief, Forrester Res. shopping. Technical Report 1.

Ström, R., M. Vendel and J. Bredican, 2014. Mobile marketing: A literature review on its value for consumers and retailers. J. Retail. Consumer Serv., 21: 1001-1012.

Teo, T.S., 2001. Demographic and motivation variables associated with Internet usage activities. Internet Res., 11: 125-137.

Venkatesh, V. and F.D. Davis, 2000. A theoretical extension of the technology acceptance model: Four longitudinal field studies. Manage. Sci., 46: 186-204.

William, A. and T. Tullis, 2013. Measuring the user experience: Collecting, analyzing and presenting usability metrics. Newnes.

Wu, J.H. and S.C. Wang, 2005. What drives mobile commerce?: An empirical evaluation of the revised technology acceptance model. Inform. Manage., 42: 719-729.

Wu, J.H. and T.L. Hisa, 2008. Developing e-business dynamic capabilities: An analysis of e-commerce innovation from I-, m-, to U-commerce. J. Organiz. Comput. Electro. Commerce Serv. Electro. Commerce Res., 12: 225-248.

YeeâLoong Chong, A. and K. Ooi, 2008. Adoption of interorganizational system standards in supply chains. Industrial Manage. Data Syst., 108: 529- 547.

Yu, J., H. Lee, I. Ha and H. Zo, 2017. User acceptance of media tablets: An empirical examination of perceived value. Telemat. Inform., 34: 206-223.

Zabadi, A.M., 2016. Adoption of Information Systems (IS): The factors that influencing is usage and its effect on employee in Jordan Telecom Sector (JTS): A conceptual integrated model. Int. J. Bus. Manage., 11: 25-25.

Zarmpou, T., V. Saprikis, A. Markos and M. Vlachopoulou, 2012. Modeling usersâ acceptance of mobile. Ergonomics Manufacturing Service Industries, Acceptance Mobile, 23: 279-293.

Zhou, L., L. Dai and D. Zhang, 2007. Online shopping acceptance model-a critical survey of consumer factors in online.
542